\documentclass[
twocolumn,
showpacs,
nofootinbib,
nobibnotes,
amsmath,amssymb,
aps,
pra,
longbibliography,
floatfix,
]{revtex4}
\usepackage{color}
\usepackage{amsmath}
\usepackage{amsxtra}
\usepackage{amstext}
\usepackage{amssymb}
\usepackage{latexsym}
\usepackage{dsfont}
\usepackage[caption=false]{subfig}
\usepackage{graphicx} 
\usepackage{float}
\usepackage{dcolumn} 
\usepackage{bm} 
\usepackage{hyperref} 
\usepackage{natbib}

\renewcommand{\vec}[1]{{\bf{#1}}}

\begin{document}
\preprint{APS/123-QED}

\title
{A tight binding model for quantum spin Hall effect on triangular optical lattice}

\author{Ahad K. Ardabili }
\affiliation{Department of Physics, Ko\c{c} University, Sar{\i}yer, Istanbul, 34450, Turkey}

\author{Tekin Dereli}
\affiliation{Department of Physics, Ko\c{c} University, Sar{\i}yer, Istanbul, 34450, Turkey}

\author{\"Ozg\"ur E. M\"ustecapl{\i}o\u{g}lu}
\affiliation{Department of Physics, Ko\c{c} University, Sar{\i}yer, Istanbul, 34450, Turkey}
\email{omustecap@ku.edu.tr}

\date{\today}

\begin{abstract}
We propose a tight binding model for the quantum spin Hall system on triangular optical lattice and we determined the edge state spectrum
which contains gap traversing states as the hallmark of $\mathds{Z}_{2}$ topological insulator. The advantage
of this system is the possibility of implementing it in the fermionic ultracold atomic system whose nearly free electron
limit is proposed by B. B\'eri and N. R. Cooper, Phys.~Rev.~Lett. {\bf 107}, 145301 (2011).
\end{abstract}



\maketitle

\section{Introduction} \label{sec:introduction}
Topological insulators (TIs) are
insulating in the bulk but have metallic states on their boundaries \cite{hasan,qi}.
Robustness of these states against disorder and perturbations makes them promising
for applications such as spintronics \cite{moore2010} and topological quantum
computation \cite{nayak}. Topological invariants of the bulk material are essential
for the robust boundary modes. This urged consideration of topological insulators
on different lattice geometries \cite{lattices1,lattices2,lattices3,lattices4,lattices5}.

It is widely acknowledged that the cold atomic systems are
ideal systems to simulate solid-state phenomena in
a controlled way. The two and three dimensional
topological insulators with band gaps in the order
of the recoil energy have recently been proposed
in ultracold fermionic atomic gases \cite{z2cooper}.
The proposal utilizes
interactions which preserves time reversal symmetry (TRS),
analogous to synthesized spin-orbit coupling  \cite{spielman},
so that the insulators are classified by the so
called $\mathds{Z}_{2}$  topological invariant \cite{fu}.

Even if the band gap in tight binding models are
 not as large as in nearly free electron
limit,  TIs in ultra cold atomic systems have been studied vastly in
tight binding regime ~\cite{imprint1,imprint2}. The optical lattices are described by continuous
potentials formed by the combinations of standing waves.
It is convenient
to treat them as deep potentials.
Our aim in this article is to propose a tight binding model for the
quantum spin Hall effect which can be realized in the ultracold atomic systems.
The corresponding model in the nearly free electron  limit is proposed by
B\'eri and Cooper ~\cite{z2cooper} with this advantage that the band gap
is large. We also determine the band structure of the edge state which
exhibits the hallmark of TIs due to its robustness against all
perturbations that preserve the TRS .

In the Sec.~\ref{sec:tight} of this paper we propose
the  tight binding model for quantum spin Hall (QSH) system
in the triangular optical lattice.  In Sec.~\ref{sec:model} we briefly
review the proposal of $\mathds{Z}_{2}$ topological
insulator in ultracold atomic gases \cite{z2cooper}.
We conclude in Sec.~\ref{sec:conclusion}.
\section{Tight Binding model}\label{sec:tight}
\subsection{Bulk band structure}

The charge quantum Hall effect depends on the breaking of time-reversal symmetry
and it has been shown that even in the absence of average non-zero external magnetic
field the quantum Hall effect can be created \cite{haldane}. However in the QSH effect
one needs to preserve the time reversal invariance. Among the first models proposed
for dissipationless QSH effect are the works by Bernevig and Zhang  \cite{bernevig}
and by Kane and Mele \cite{kane}, where the authors used the spin-orbit coupling
such that the two different spin direction  experiences the same magnetic field strength but
with opposite sign. In other words their system were two copies of a quantum Hall system
for each spin where the total first Chern number adds up to zero and the system is
time reversal invariant.

\begin{figure}[t]
\centering
\includegraphics[width=0.3\textwidth]{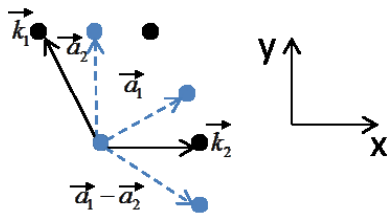}
\caption{\label{fig:fig1}
 (Color online)  The light points and dashed lines
show the lattice sites of the atoms and the hopping directions in
 the triangular lattice respectively, and the dark
points and lines  are the corresponding Brillouin zone sites and vectors respectively. }
\end{figure}

Physically our model corresponds to the same scenario.
We propose a Hamiltonian for a fermion on triangular lattice Fig.~\ref{fig:fig1}
with a mirror symmetric spin orbit coupling as,
\begin{align}\label{hamiltonian2d}
   H &= t\sum_{m,n} C^{\dag}_{m+1,n} C_{m,n} +  C^{\dag}_{m,n+1}e^{i 4 \pi m \phi}\sigma_{z} C_{m,n}\nonumber\\
     &+  C^{\dag}_{m+1,n-1} e^{i 2(m+1)\pi \phi} \sigma_{z} C_{m,n},
\end{align}
where $\phi = p/q$ is flux per plaquette and we take $p = 1$ and $q = 4$ in this paper.
$C_{m,n} = (c_{m,n\uparrow},c_{m,n\downarrow})^{T} $ and $C^{\dag}_{m,n} $ are
annihilation and creation operators
on site $(m,n)$ respectively.   We take the hopping parameter $t = 1$ throughout this paper.
The first term is nearest neighbour hopping term on the triangular lattice
with $\mathbf{a}_{1} = (\sqrt{3}/4,1/4)a$ and $\mathbf{a}_{2} = (0,1/2)a$, where $a$ is
the lattice constant (see Fig. ~\ref{fig:fig1}). The second and third terms are
mirror symmetric spin-orbit interaction. $\sigma_{z}$ is
the Pauli matrix. In the absence of spin
this Hamiltonian implies that electron acquires $\phi=1/4$
of flux quantum enclosing the elementary plaquette of the
triangular lattice.

In order to calculate the band structure
we take the Fourier transform of the Hamiltonian Eq.~(\ref{hamiltonian2d}).
We use the momentum representation of fermionic operator
\begin{equation}
C_{\mathbf{k}} =  \sum_{m,n} e^{i \mathbf{k}\cdot \mathbf{R}_{m,n} }C_{m,n},
\end{equation}
where  $\mathbf{R}_{m,n} = m\vec{a}_{1}+n\vec{a}_{2}$. We obtain the energy dispersion in triangular lattice
by solving the determinant for the eigenvalues $\epsilon$,
\begin{equation}\label{Dispersion}
   \mbox{Det} \left[
    \begin{array}{cccc}
      - A-\epsilon & B + i C& 0 & 0 \\
      B - i C  &  A-\epsilon & 0 & 0\\
           0 & 0 & A-\epsilon & B - i C \\
           0 & 0 & B + i C  &  -A-\epsilon \\
    \end{array}
    \right] = 0,
\end{equation}
where $A$,$B$ and $C$ are defined to be,
\begin{align}
    A &=  \cos\left(k_{y}a/2\right), \\
    B &=  \cos\frac{a}{4}\left(\sqrt{3}k_{x}+k_{y}\right),\\
    C &= \cos\frac{a}{4}\left(\sqrt{3}k_{x}-k_{y}\right).
\end{align}
\begin{figure}[t]
\subfloat[][]{
\includegraphics[width=0.4\textwidth]{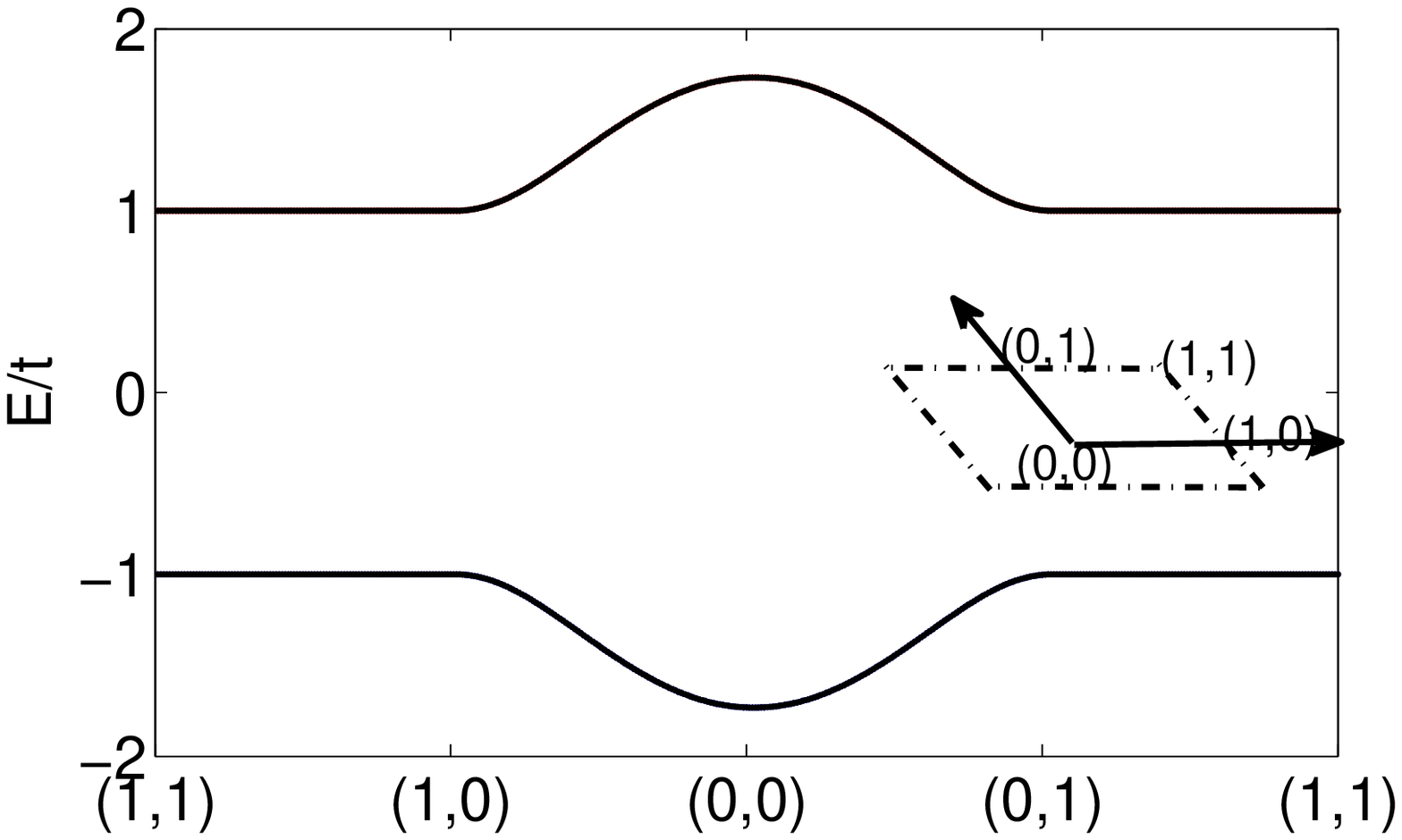}
\label{fig:fig2a}}
\qquad
\subfloat[][]{
\includegraphics[width=0.40\textwidth]{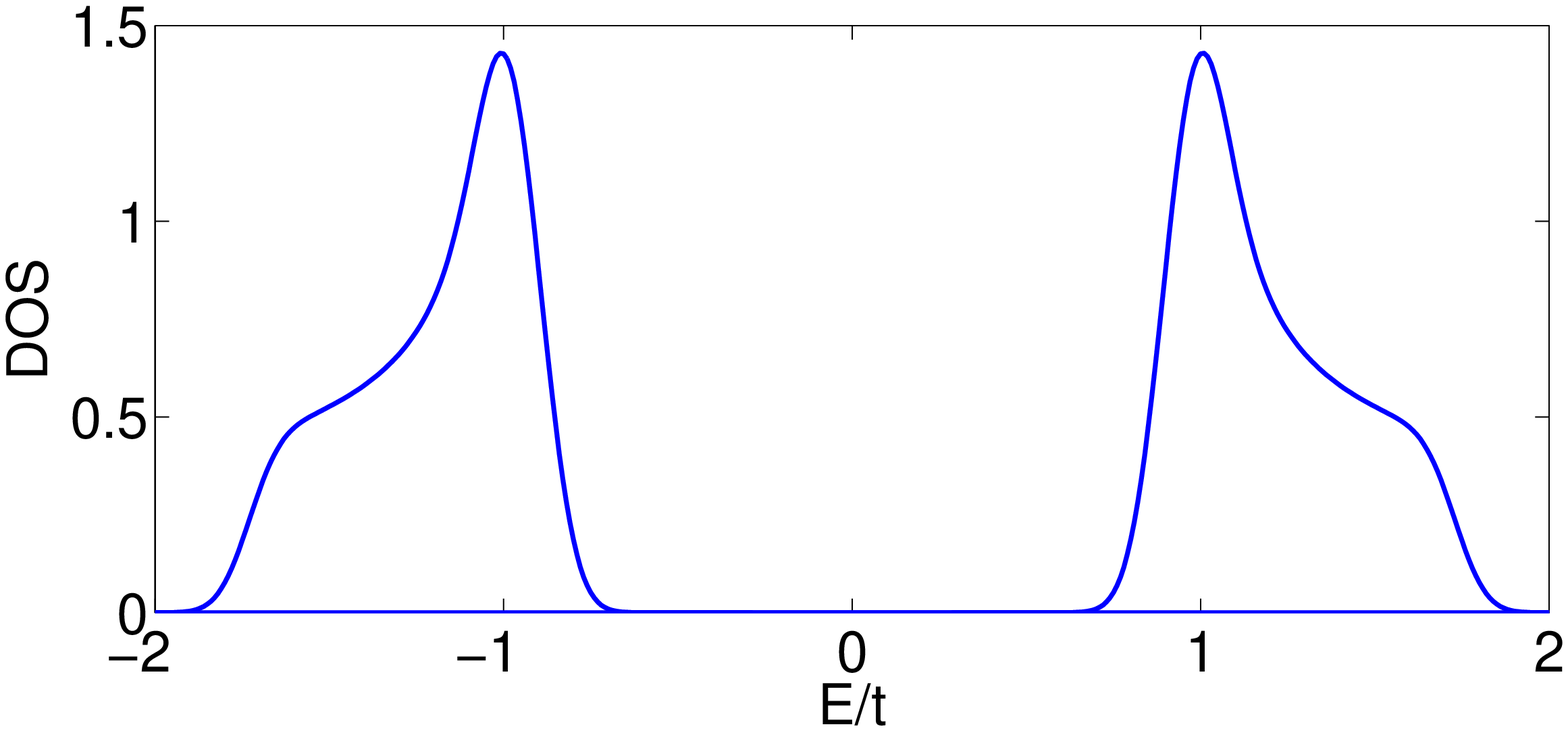}
\label{fig:fig2b}}
\qquad
\subfloat[][]{
\includegraphics[width=0.4\textwidth]{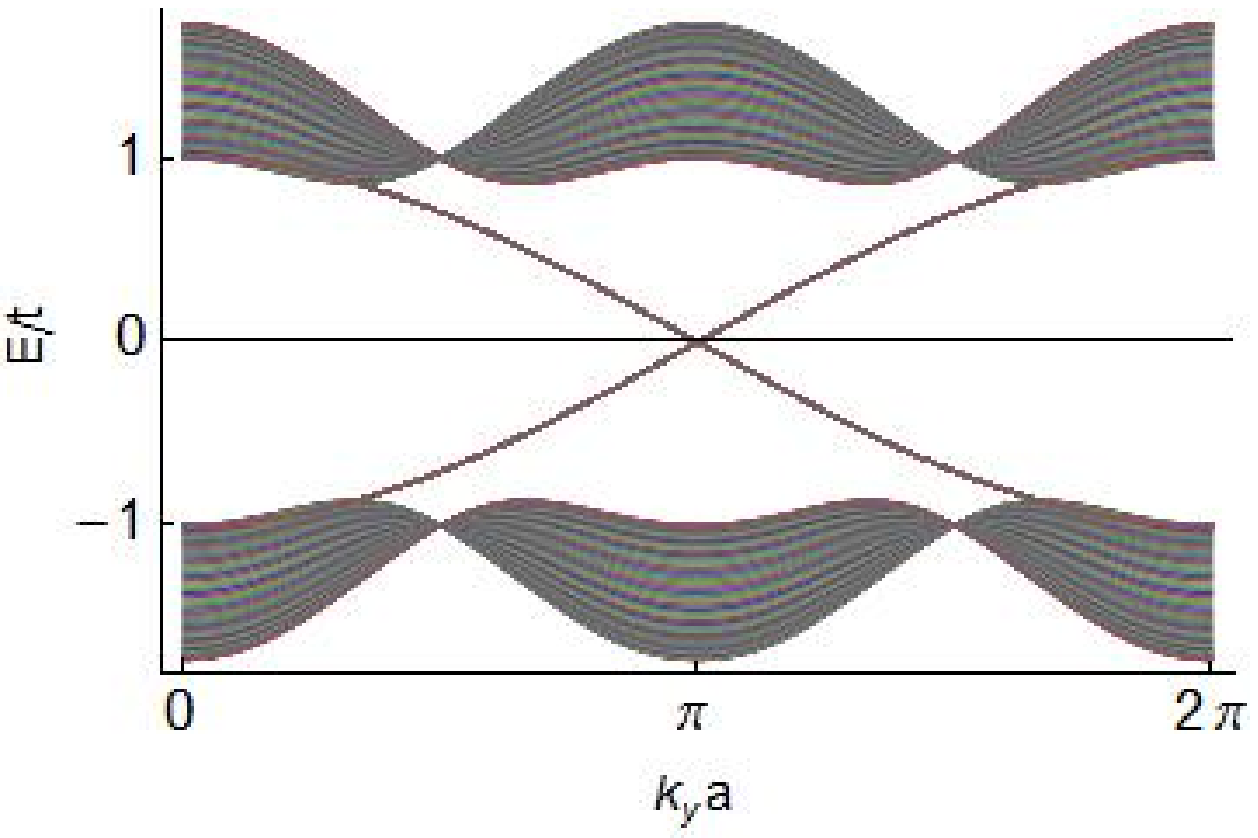}
\label{fig:fig2c}}
\caption{
\label{fig:fig2}
(Color online)
(a) The band structure of the Hamiltonian Eq.
(\ref{hamiltonian2d}) along the  path (1,1), (1,0), (0,0), (0,1) where 1 and 0
are referring to the inset (m,n) in $\vec{\Gamma}_{mn}=(m \vec{k}_{1}+n \vec{k}_{2})$,
which are the TRS invariant points in Brillouin zone. The bands are  four fold degenerate
as in the nearly free electron limit of cold atomic system.
(b) The DOS for all the energy bands is shown where the horizontal axis
 is the energy.
(c)  The spectrum of the edges of triangular lattice.
 The shaded area is  lowest energy bands for the triangular
lattice (for uniform spacing of $ -4 \pi/\sqrt{3}< k_{x}a < 4\pi/\sqrt{3}$).
  The solid lines show the spin polarized
 edge state band, traversing
 the gap. The edge states cross at $k_{y}a=\pi$.  }
\end{figure}
In order to solve Eq. (\ref{Dispersion}), we used a 2D grid for the
$\vec{k}$-space.  Fig.~ \ref{fig:fig2a} shows the band structure of
the Eq.~ (\ref{Dispersion}) for a cell with specific $\vec{k}$ points as
its corners taken to be $(k_{1}+k_{2})/2$, $k_{1}/2$, $0$, $k_{2}/2$, as shown in the inset.
These points are the TRS invariant points in the Brillouin zone
Since each of the two blocks of
the Eq.~ (\ref{Dispersion}) corresponds to two fold spin degenerate
bands, each band of the Fig.~ \ref{fig:fig2a}
is four fold degenerate.

We also calculated the density of states (DOS) for the Hamiltonian in Eq.~(\ref{Dispersion}).
This quantity is defined by the expression,
\begin{equation}\label{dos}
    \rho(E) = \frac{1}{A} \sum_{\mathbf{k}} \delta(E-E_{n}),
\end{equation}
where $A$ is the area of the system in reciprocal space and $E_{n}$ is the energy of the bands.
In Fig. \ref{fig:fig2b} the DOS is depicted for all the energy bands.

\subsection{Edge-state band structure}
The characteristic of the $\mathds{Z}_{2}$ topological insulator
is the gapless edge states. They describe two spin currents at the edge, propagating
in opposite direction. This property
is because of the the time-reversal symmetry and it prevents the gap
opening due to any TR invariant perturbation as the
result of the Kramer's theorem ~\cite{kane}.

 We follow the method in Ref.~\cite{hatsugai} to find the energy dispersion of the
edge states. The Hamiltonian Eq.~(\ref{hamiltonian2d}),
must be reduced to a one-dimensional problem.
We take the {\it y} direction as the
periodic part and we use the momentum representation as,
\begin{equation}
  C_{m,n} = \frac{1}{\sqrt{L_{y}}}\sum_{k_{y}}e^{i k_{y} n}C_{m}(k_{y}),
\end{equation}
where $k_{y} a/2  = 2\pi n_{y}/L_{y}$, $n_{y} = 1,...,L_{y}$ and $L_{y}$ is the system size
along $y$ direction. By inserting the
 single particle state
\begin{equation}
 |\Psi (k_{y})\rangle = \sum _{m}|\Psi (k_{y})_{m}\rangle C^{\dag}_{m}(k_{y})|0\rangle
\end{equation}
into the Schr\"{o}dinger equation $H|\Psi\rangle = E|\Psi\rangle$, the spin up part of the
problem is reduced to the one-dimensional problem  with parameter
$k_{y}$ as
\begin{eqnarray}\label{harper}
  G^{*} \Psi_{m+1} -  G\Psi_{m-1} - 2 \cos \left( k_{y}a/2 - 4 \pi \phi m \right)\Psi_{m} \nonumber \\
   =  E \psi_{m}.& &
\end{eqnarray}
where $G = 1+e^{-i\left( -k_{y}a/2+\pi \phi\left(2m+1\right)\right)}$. Including the spin down as well,
this equation can be written
as a generalized Harper equation~\cite{harper} in transfer matrix form,
\begin{equation}\label{transfer matrix}
   \left(
     \begin{array}{c}
       \Psi_{m+1\uparrow}\left(k_{y}\right) \\
       \Psi_{m\uparrow}\left(k_{y}\right) \\
       \Psi_{m+1\downarrow}\left(k_{y}\right) \\
       \Psi_{m\downarrow}\left(k_{y}\right) \\
     \end{array}
   \right) = M
   \left(
     \begin{array}{c}
       \psi_{m\uparrow}\left(k_{y}\right) \\
       \psi_{m-1\uparrow}\left(k_{y}\right) \\
       \psi_{m\downarrow}\left(k_{y}\right) \\
       \psi_{m-1\downarrow}\left(k_{y}\right) \\
     \end{array}
   \right),
\end{equation}
where $M$ is the transfer matrix, which is given by:
%
\begin{eqnarray}\label{harper}
 M = \left(
       \begin{array}{cc}
         \begin{array}{cc}
           \frac{F}{G} & \frac{G^{*}}{G} \\
           1 & 0
         \end{array}
          & 0 \\
         0 & \begin{array}{cc}
           \frac{F'}{G'} & \frac{G'^{*}}{G'} \\
           1 & 0
         \end{array} \\
       \end{array}
     \right)
\end{eqnarray}
with $G' = 1-e^{-i\left( -k_{y}a/2+\pi \phi\left(2m+1\right)\right)}$,
 $F = -\epsilon-2\cos \left(k_{y}a/2-4\pi\phi m\right)$ and
$F' = -\epsilon+2\cos \left(k_{y}a/2-4\pi\phi m\right)$.
Under the boundary condition that the wavefunction
goes to zero at the boundaries of lattice we can solve this equation.
The band structure along the path $k_{y}a=0-2\pi$ is shown in the Fig.~ \ref{fig:fig2c}.
 We used 100 k-points along $k_{y}$ direction.
The shaded area is the bulk band and the gap traversing edge states as the signature of  $\mathds{Z}_{2}$
topological insulator are plotted as the solid line. Since the TRS is preserved
no TR symmetric perturbation can open the gap at $k_{y}a=\pi$ \cite{kaneQSH}.

\section{Cold Atomic system} \label{sec:model}
In this section we review briefly the $\mathds{Z}_{2}$ topological insulator
model proposed by B\'eri and cooper~\cite{z2cooper}. This model is studied in
nearly free electron limit which has the advantage of large band gap.
\begin{figure}[t]
\includegraphics[width=0.3\textwidth]{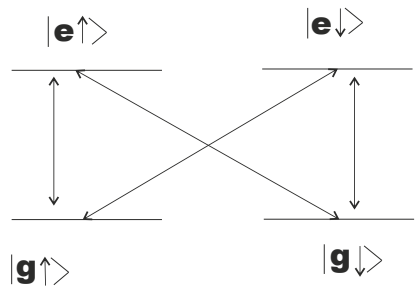}
\caption{
\label{fig:fig3}
(Color online)
 The atomic levels of Ytterbium and the interactions between atoms and lasers }
\end{figure}

The Hamiltonian  which describes an atom with position
$\vec{r}$ and momentum $\vec{p}$ and with N internal
states  is given by

\begin{equation}\label{hamiltonianColdatom}
H=\frac{\vec{p}^{2}}{2m}\mathds{1}_{4} + V\hat{M}(\vec{r}),
\end{equation}
where $V\hat{M}(\vec{r})$ is a position dependent potential.
In order to have a system with low spontaneous emission
one can use  ytterbium (Yb) which has long-lived excited state.
The two internal states, ground state ($^{1}S_{0}=g$) and long-lived
excited state ($^3p_{0}=e$) of Yb have spin degree of freedom
which leads to four states Fig. ~\ref{fig:fig3}. Another interesting aspect of Yb is the
existence of  a state dependent scalar potential for $\lambda_{\text{magic}}$
with opposite sign $\pm V_{\text{am}}(\vec{r})$. Therefore we can write the
potential part of hamiltonian when we have external electric field as
$\vec{E}=\vec{\epsilon}e^{-i\omega t}+\vec{\epsilon}^{*}e^{i\omega t}$ with complex amplitude $\epsilon$
and frequency $\omega$.
 All four e-g transitions have the same frequency $\omega_{0}=(E_{e}-E_{g})/\hbar$.
Using rotating wave approximation \cite{1} we have the optical potential as following,
\begin{equation}\label{coupling}
V\hat{M}(\vec{r}) = \left( \begin{array}{ccc}
 (\frac{\hbar}{2}\Delta+V_{\text{am}})1& -i\vec{\sigma}.\vec{\epsilon}d_{r}  \\
i\vec{\sigma}.\vec{\epsilon}d_{r} & -(\frac{\hbar}{2}\Delta+V_{\text{am}}) \end{array} \right)\
\end{equation}
where $\Delta=\omega-\omega_{0}$ is the atom-field  detuning and $d_{r}$ is the dipole moment.
One can write the hamiltonian in terms of
Dirac matrices ~\cite{2},
\begin{equation}
\Gamma^{1, 2, 3} = -i\sigma^{y}\otimes \sigma^{i} ,\Gamma^{4}= \sigma^{x}\otimes I ~\text{and}~ \Gamma^{5}=\sigma^{z}\otimes I,
\end{equation}
which gives,
\begin{equation}\label{eq:hamil}
   H = \frac{\hat{\mathbf{P}}^{2}}{2m}\mathds{1} +\Gamma^{i}d_{r}\epsilon_{i}+\Gamma^{5}(\frac{\hbar}{2} \Delta+V_{\text{am}}).
\end{equation}

For the two-dimensional system one can make following choice for the potential matrix Eq.~(\ref{coupling}):
\begin{eqnarray}
  d_{r} \mathbf{\epsilon}=[V\delta,V\cos(\vec{r}.\vec{k}_{1}), V\cos(\vec{r}.\vec{k}_{2})],\nonumber\\
\frac{\hbar}{2}\Delta+V_{am}(\vec{r})=V\cos(\vec{r}.(\vec{k}_{1}+\vec{k}_{2}))\label{eq:elements}
\end{eqnarray}
with $\vec{k}_{1}=k(1,0,0)$ and $\vec{k}_{2}=k(\cos(\theta),\sin(\theta),0)$.
\begin{figure}[t]
\subfloat[][]{
\includegraphics[width=0.4\textwidth]{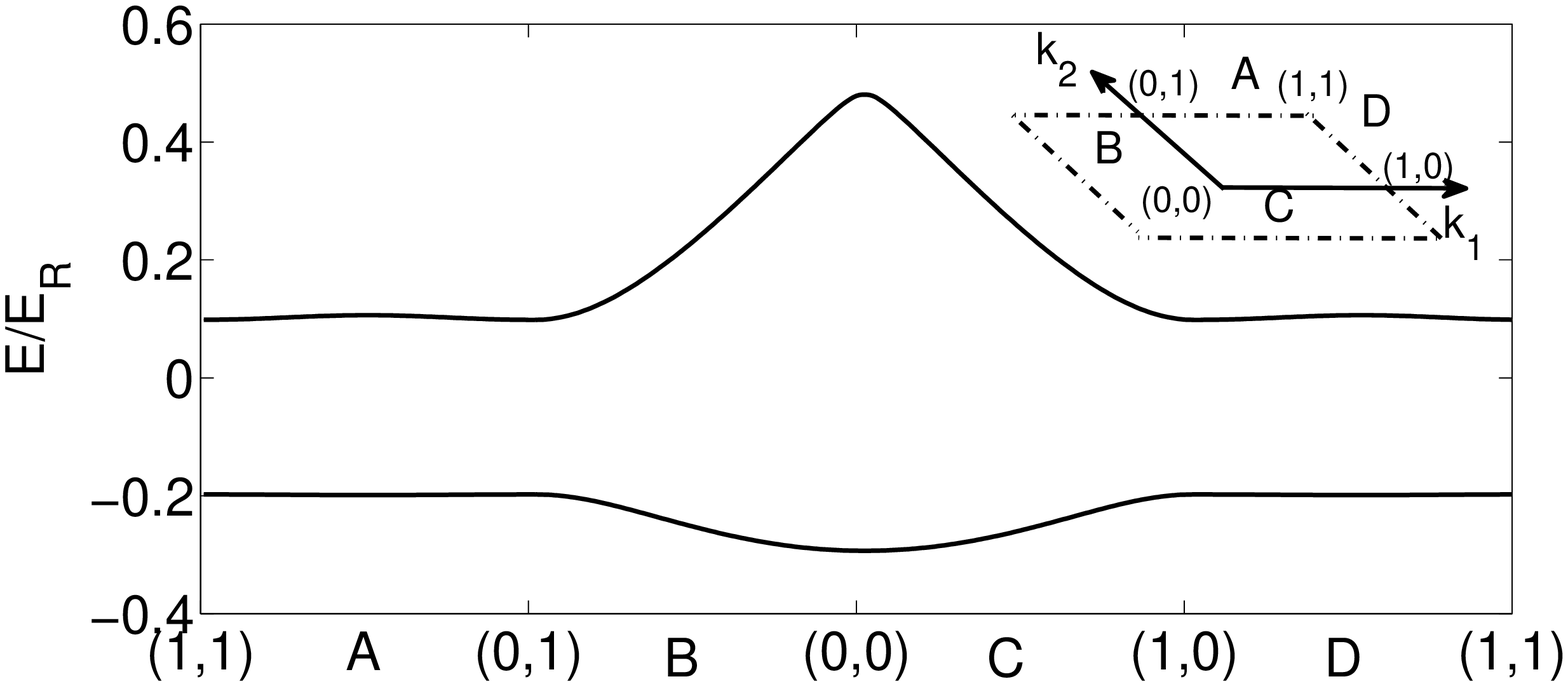}
\label{fig:fig4a}}
\qquad
\subfloat[][]{
\includegraphics[width=0.4\textwidth]{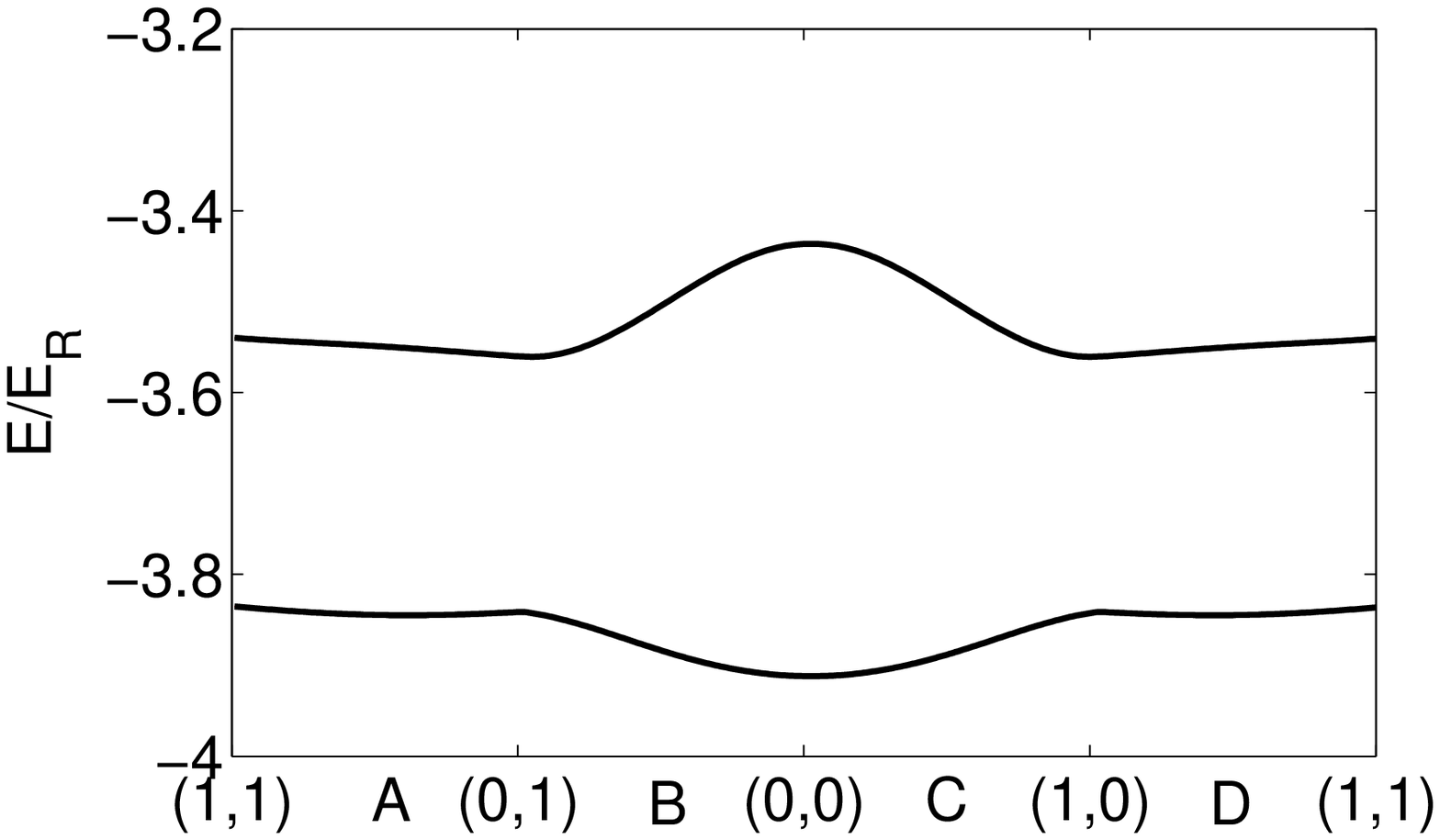}
\label{fig:fig4b}}
\qquad
\subfloat[][]{
\includegraphics[width=0.4\textwidth]{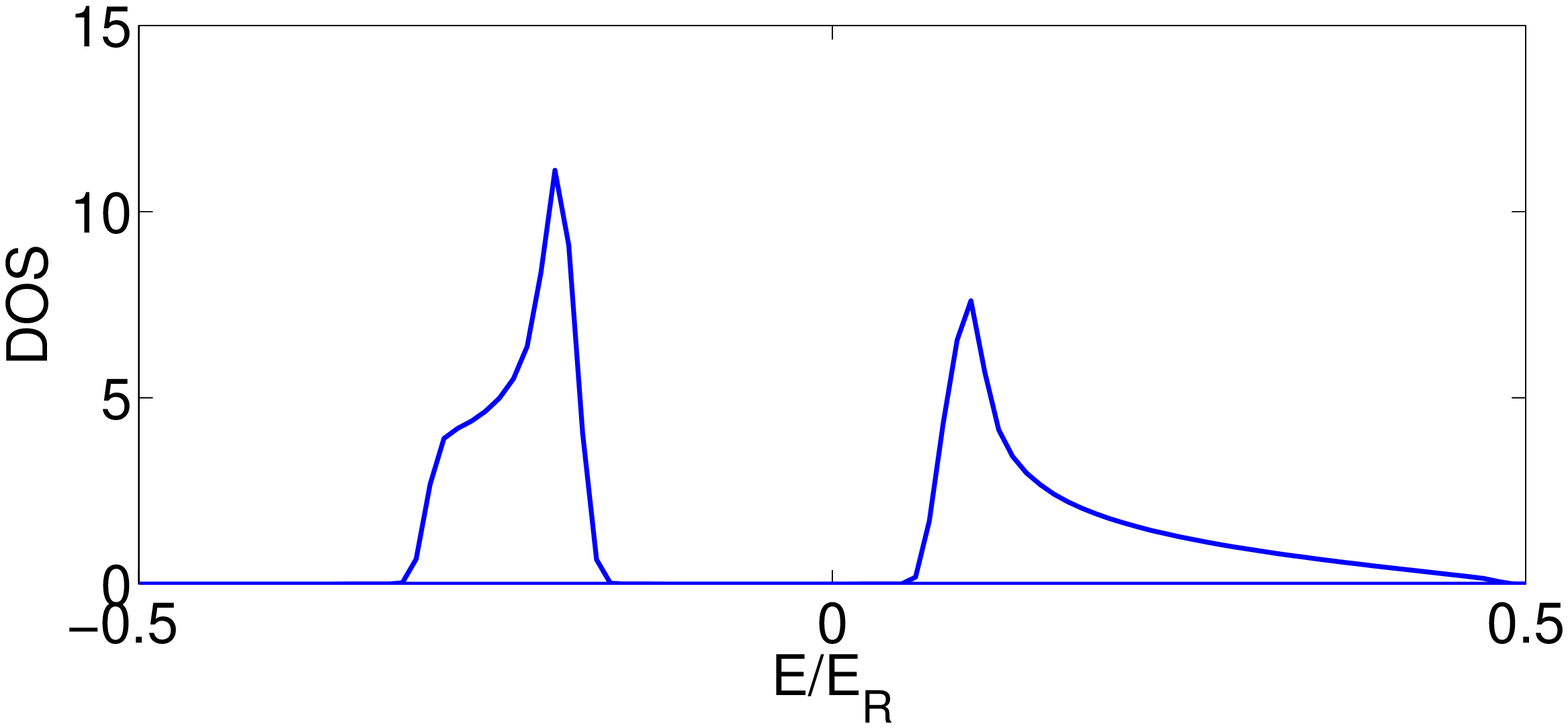}
\label{fig:fig4c}}
\caption{
\label{fig:fig4}
(Color online)
 (a) Lowest energy bands of the cold atom system in nearly free electron limit as the result
 of the solution of Eq.~(\ref{eq:hamil}). The energy is plotted relative
 to recoil energy $E_{R}$. $V=0.5 E_{R}$ and $\delta=0$ are shown here.
The k-points are labeled as $\vec{\Gamma}_{mn}=(m \vec{k}_{1}+n \vec{k}_{2})/2$ and
each band is four fold degenerate. (b) Energy bands for the potential $V=3.5 E_{R}$ which
resembles the band structure of tight binding regime Fig.~\ref{fig:fig2a}. (c) DOS of the energy bands
in Fig.~\ref{fig:fig2a} which are depicted
in the band structure is shown. Energy is expressed in the unit of $E_{R}$}
\end{figure}
The optical potential in Eq.~(\ref{Adiabaticpotential}) is formed from three standing
waves which are linear polarized light at the coupling
frequency $\omega$. Two of these waves have equal amplitude
in the 2D plane ($\vec{k}_{1}$ for y polarization and $\vec{k}_{2}$ for $\vec{z}$ polarization)
the $\vec{x}$ polarized wave vector is normal to the $2$D plane
with an amplitude smaller by a factor of $\delta$.
Since the $\omega\simeq\omega_{0}$ , we have $k\simeq2 \pi / \lambda_{0}$ with
$\lambda_{0}=578~ \text{nm}$ the wavelength of the e-g transition.
The spatial dependence of $V_{\text{am}}$ is set by a standing
wave at the antimagic wavelength $\lambda_{\text{am}}$ \cite{dalibard},
which creates a state-dependent potential with $ |\vec{k}_{1}+\vec{k}_{2}|=4 \pi / \lambda_{\text{am}}$
that leads  $\theta=2 \arccos(\pm\lambda_{0}/\lambda_{\text{am}})$. For simplicity,
in all following discussions one can fix $\theta=2\pi/3$ and define $a\equiv4\pi/(\sqrt{3}k)$.
Therefore the optical coupling $\hat{M}$ has the symmetry of a triangular lattice.
In Fig.~\ref{fig:fig4a} we show the few
lowest energy bands for $\delta= 0$. The bands
were calculated by numerical diagonalization in the plane
wave basis (49 plane waves is used). All bands are fourfold degenerate similar to
tight binding regime.

The relation of this system to the tight binding model
given in previous section becomes more clear as one applies the
unitary transformation ~$\hat{U} = \left(\mathds{1}-i \hat{\Sigma}_{3} \hat{\sigma}_{2}\right)/\sqrt{2}$
to the coupling  $\hat{M}$ in Eq. (\ref{coupling}):
\begin{equation}\label{unitary potential}
  \hat{M'} = \hat{U}^{\dag}\hat{M} \hat{U} = c_{1}\hat{\Sigma}_{1}+
  c_{2}\hat{\Sigma}_{2}\sigma_{3} + c_{12} \hat{\Sigma}_{3} ,
\end{equation}
\begin{figure}[t]
\centering
\includegraphics[width=0.4\textwidth]{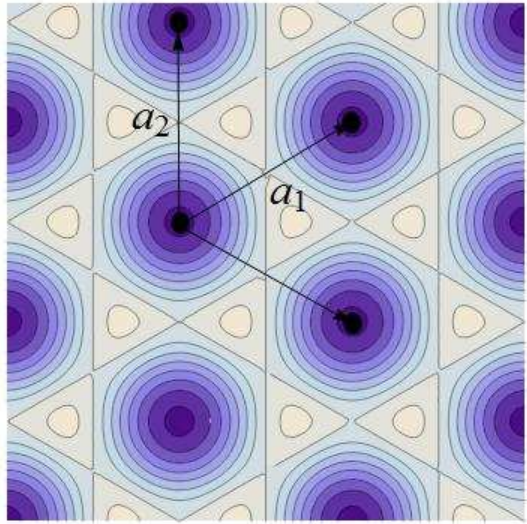}
\caption{\label{fig:fig5}
 (Color online)  The dark circles
show the local minima of the adiabatic energy which forms
a triangular lattice in the tight binding limit. }
\end{figure}
here $\Sigma_{i}= \sigma_{i}\otimes \mathds{1}_{2\times 2}$,
$c_{i} \equiv \cos (\mathbf{r} \cdot \mathbf{k_{i}} ) $ and $c_{12}=
\cos \left(\mathbf{r}\cdot \left(\mathbf{k}_{1}+\mathbf{k}_{2}\right)\right)$. This matrix is
$2\times 2$ block diagonal matrix for each eigenvalue of $\sigma_{3}$
( since the kinetic part is diagonal this is the case for the Hamiltonian as well)
thus the four level system decouples into
two two  level system each of which experiences an effective magnetic field
due to the optical dressed state of the $c_{1}\hat{\Sigma}_{1}+
\pm c_{2}\hat{\Sigma}_{2} + c_{12} \hat{\Sigma}_{3}$ ~\cite{opticalflux}. This means that
opposite spin direction undergoes an effective magnetic field of the same strength
but with opposite signs.  Beside the lowest band energy of these systems
have $\pm 1$ Chern number which they cancel out each other because of the
time-reversal symmetry. These are the required criteria for the
quantum spin Hall effect.

 To make the connection to the previous section
we consider the adiabatic limit $V \gg \hbar k^{2}/2M $~\cite{gauge} when the
potential part of the Hamiltonian Eq.~(\ref{hamiltonianColdatom}) plays the dominant
role. In order to find the minima of the adiabatic energy which gives the lattice
sites in tight binding regime ~\cite{toolbox} we diagonalized the  potential in Eq. ~(\ref{unitary potential})
analytically and obtained,
 \begin{equation}\label{Adiabaticpotential}
   V_{\text{ad}} = \pm \sqrt{c_{1}^{2}+c_{2}^{2}+c_{12}^{2}},
 \end{equation}
as plotted in Fig.~\ref{fig:fig5}.
Now if we ignore the spin here,
the effective magnetic field strength experienced by the
neutral atom following the adiabatic path is equivalent to the
$1/4$ of a flux quantum that a charged particle
acquires enclosing the elementary plaquette of the triangular lattice in
tight binding limit ~\cite{opticalflux}. This is  equivalent to the
Hamiltonian Eq. (\ref{hamiltonian2d}) proposed in this paper without spin.
Therefore tight binding limit of the ultra cold atomic system of Ref.~\cite{z2cooper}
is given in Eq. (\ref{hamiltonian2d}) and the hopping parameter $t$ is related to the potential
scale of optical coupling $V$ based on the formalism in \cite{zoller}.
The difference is in the $(0,0)$ point in k-space where the upper band
of cold atom limit has a sharper peak Fig.~\ref{fig:fig4a} than the tight binding upper band Fig~\ref{fig:fig2a}.
This can be understood as the characteristic behavior of the energy levels of free electron
 which are just a  parabola in k (momentum), by getting distorted due to a periodic potential \cite{ashcroft}. As the potential
becomes stronger the energy dispersion resembles the tight binding regime Fig.~\ref{fig:fig4b}.
The DOS for the nearly free electron limit is also depicted in Fig.~\ref{fig:fig4c} which shows that
states are distributed around the two energy bands across the gap asymmetrically due to the asymmetrically
 located van Hove singularities of the upper and lower bands in contrast to that of tight binding case in Fig.~\ref{fig:fig2b}.
 Finally we note that realization of QSH models in Eq.~(\ref{hamiltonian2d}) and Eq. (\ref{hamiltonianColdatom}) does not
 require any spin flipping interactions and thus does not need any additional cooling mechanism \cite{spinflip}.
\section{Conclusion}\label{sec:conclusion}
Summarising, we considered the quantum spin Hall effect on the
triangular lattice in the tight binding limit and we proposed that
this can be realized in the ultra cold atomic system. We studied
the edge state band structure which reveals the $\mathds{Z}_{2}$.  The nearly free electron limit
of the system we proposed here is introduced  as $\mathds{Z}_{2}$ topological
insulator in Ref.~\cite{z2cooper}.
\acknowledgements
Authors acknowledge useful discussions with O. Oktel and I. Adagideli.
\"O.E.M. and A. K. A. acknowledge TUBITAK Project No. 112T974 for support.

\begin{thebibliography}{29}
\expandafter\ifx\csname natexlab\endcsname\relax\def\natexlab#1{#1}\fi
\expandafter\ifx\csname bibnamefont\endcsname\relax
  \def\bibnamefont#1{#1}\fi
\expandafter\ifx\csname bibfnamefont\endcsname\relax
  \def\bibfnamefont#1{#1}\fi
\expandafter\ifx\csname citenamefont\endcsname\relax
  \def\citenamefont#1{#1}\fi
\expandafter\ifx\csname url\endcsname\relax
  \def\url#1{\texttt{#1}}\fi
\expandafter\ifx\csname urlprefix\endcsname\relax\def\urlprefix{URL }\fi
\providecommand{\bibinfo}[2]{#2}
\providecommand{\eprint}[2][]{\url{#2}}

\bibitem[{\citenamefont{Hasan and Kane}(2010)}]{hasan}
\bibinfo{author}{\bibfnamefont{M.~Z.} \bibnamefont{Hasan}} \bibnamefont{and}
  \bibinfo{author}{\bibfnamefont{C.~L.} \bibnamefont{Kane}},
  \bibinfo{journal}{Rev. Mod. Phys.} \textbf{\bibinfo{volume}{82}},
  \bibinfo{pages}{3045} (\bibinfo{year}{2010}).

\bibitem[{\citenamefont{Zhang}(2011)}]{qi}
\bibinfo{author}{\bibfnamefont{X.~L.~Q. S.~C.} \bibnamefont{Zhang}},
  \bibinfo{journal}{Rev. Mod. Phys.} \textbf{\bibinfo{volume}{83}},
  \bibinfo{pages}{1057} (\bibinfo{year}{2011}).

\bibitem[{\citenamefont{Moore}(2010)}]{moore2010}
\bibinfo{author}{\bibfnamefont{J.~E.} \bibnamefont{Moore}},
  \bibinfo{journal}{Nature} \textbf{\bibinfo{volume}{464}},
  \bibinfo{pages}{194} (\bibinfo{year}{2010}).

\bibitem[{\citenamefont{Nayak et~al.}(2008)\citenamefont{Nayak, Simon, Stern,
  Freedman, , and Sarma}}]{nayak}
\bibinfo{author}{\bibfnamefont{C.}~\bibnamefont{Nayak}},
  \bibinfo{author}{\bibfnamefont{S.~H.} \bibnamefont{Simon}},
  \bibinfo{author}{\bibfnamefont{A.}~\bibnamefont{Stern}},
  \bibinfo{author}{\bibfnamefont{M.}~\bibnamefont{Freedman}}, ,
  \bibnamefont{and} \bibinfo{author}{\bibfnamefont{S.~D.} \bibnamefont{Sarma}},
  \bibinfo{journal}{Rev. Mod. Phys.} \textbf{\bibinfo{volume}{80}},
  \bibinfo{pages}{1083} (\bibinfo{year}{2008}).

\bibitem[{\citenamefont{Hu et~al.}(2011)\citenamefont{Hu, Kargarian, and
  Fiete}}]{lattices1}
\bibinfo{author}{\bibfnamefont{X.}~\bibnamefont{Hu}},
  \bibinfo{author}{\bibfnamefont{M.}~\bibnamefont{Kargarian}},
  \bibnamefont{and} \bibinfo{author}{\bibfnamefont{G.~A.} \bibnamefont{Fiete}},
  \bibinfo{journal}{Phys. Rev. B} \textbf{\bibinfo{volume}{84}},
  \bibinfo{pages}{155116} (\bibinfo{year}{2011}).

\bibitem[{\citenamefont{Weeks and Franz}(2010)}]{lattices2}
\bibinfo{author}{\bibfnamefont{C.}~\bibnamefont{Weeks}} \bibnamefont{and}
  \bibinfo{author}{\bibfnamefont{M.}~\bibnamefont{Franz}},
  \bibinfo{journal}{Phys. Rev. B} \textbf{\bibinfo{volume}{82}},
  \bibinfo{pages}{085310} (\bibinfo{year}{2010}).

\bibitem[{\citenamefont{Guo and Franz}(2009{\natexlab{a}})}]{lattices3}
\bibinfo{author}{\bibfnamefont{H.-M.} \bibnamefont{Guo}} \bibnamefont{and}
  \bibinfo{author}{\bibfnamefont{M.}~\bibnamefont{Franz}},
  \bibinfo{journal}{Phys. Rev. B} \textbf{\bibinfo{volume}{80}},
  \bibinfo{pages}{113102} (\bibinfo{year}{2009}{\natexlab{a}}).

\bibitem[{\citenamefont{Guo and Franz}(2009{\natexlab{b}})}]{lattices4}
\bibinfo{author}{\bibfnamefont{H.-M.} \bibnamefont{Guo}} \bibnamefont{and}
  \bibinfo{author}{\bibfnamefont{M.}~\bibnamefont{Franz}},
  \bibinfo{journal}{Phys. Rev. Lett.} \textbf{\bibinfo{volume}{103}},
  \bibinfo{pages}{20680} (\bibinfo{year}{2009}{\natexlab{b}}).

\bibitem[{\citenamefont{L.~Fu and Mele}(2007)}]{lattices5}
\bibinfo{author}{\bibfnamefont{C.~L.~K.} \bibnamefont{L.~Fu}} \bibnamefont{and}
  \bibinfo{author}{\bibfnamefont{E.~J.} \bibnamefont{Mele}},
  \bibinfo{journal}{Phys. Rev. Lett.} \textbf{\bibinfo{volume}{98}},
  \bibinfo{pages}{106803} (\bibinfo{year}{2007}).

\bibitem[{\citenamefont{B\'eri and Cooper}(2011)}]{z2cooper}
\bibinfo{author}{\bibfnamefont{B.}~\bibnamefont{B\'eri}} \bibnamefont{and}
  \bibinfo{author}{\bibfnamefont{N.~R.} \bibnamefont{Cooper}},
  \bibinfo{journal}{Phys. Rev. Lett.} \textbf{\bibinfo{volume}{107}},
  \bibinfo{pages}{145301} (\bibinfo{year}{2011}).

\bibitem[{\citenamefont{Lin et~al.}(2011)\citenamefont{Lin, Jim\'enez-Garc\'ia,
  and Spielman}}]{spielman}
\bibinfo{author}{\bibfnamefont{Y.-J.} \bibnamefont{Lin}},
  \bibinfo{author}{\bibfnamefont{K.}~\bibnamefont{Jim\'enez-Garc\'ia}},
  \bibnamefont{and} \bibinfo{author}{\bibfnamefont{I.~B.}
  \bibnamefont{Spielman}}, \bibinfo{journal}{Nature}
  \textbf{\bibinfo{volume}{471}}, \bibinfo{pages}{83} (\bibinfo{year}{2011}).

\bibitem[{\citenamefont{Fu and Kane}(2006)}]{fu}
\bibinfo{author}{\bibfnamefont{L.}~\bibnamefont{Fu}} \bibnamefont{and}
  \bibinfo{author}{\bibfnamefont{C.~L.} \bibnamefont{Kane}},
  \bibinfo{journal}{Phys. Rev. B} \textbf{\bibinfo{volume}{74}},
  \bibinfo{pages}{195312} (\bibinfo{year}{2006}).

\bibitem[{\citenamefont{Juzelians et~al.}(2010)\citenamefont{Juzelians,
  Ruseckas, and Dalibard}}]{imprint1}
\bibinfo{author}{\bibfnamefont{G.}~\bibnamefont{Juzelians}},
  \bibinfo{author}{\bibfnamefont{J.}~\bibnamefont{Ruseckas}}, \bibnamefont{and}
  \bibinfo{author}{\bibfnamefont{J.}~\bibnamefont{Dalibard}},
  \bibinfo{journal}{Phys. Rev. A} \textbf{\bibinfo{volume}{81}},
  \bibinfo{pages}{053403} (\bibinfo{year}{2010}).

\bibitem[{\citenamefont{Goldman et~al.}(2010)\citenamefont{Goldman, Satija,
  Nikolic, Bermudez, Martin-Delgado, Lewenstein, and Spielman}}]{imprint2}
\bibinfo{author}{\bibfnamefont{N.}~\bibnamefont{Goldman}},
  \bibinfo{author}{\bibfnamefont{I.}~\bibnamefont{Satija}},
  \bibinfo{author}{\bibfnamefont{P.}~\bibnamefont{Nikolic}},
  \bibinfo{author}{\bibfnamefont{A.}~\bibnamefont{Bermudez}},
  \bibinfo{author}{\bibfnamefont{M.~A.} \bibnamefont{Martin-Delgado}},
  \bibinfo{author}{\bibfnamefont{M.}~\bibnamefont{Lewenstein}},
  \bibnamefont{and} \bibinfo{author}{\bibfnamefont{I.~B.}
  \bibnamefont{Spielman}}, \bibinfo{journal}{Phys. Rev. Lett.}
  \textbf{\bibinfo{volume}{105}}, \bibinfo{pages}{255302}
  (\bibinfo{year}{2010}).

\bibitem[{\citenamefont{Haldane}(1988)}]{haldane}
\bibinfo{author}{\bibfnamefont{F.~D.~M.} \bibnamefont{Haldane}},
  \bibinfo{journal}{Phys. Rev. lett} \textbf{\bibinfo{volume}{61}},
  \bibinfo{pages}{2015} (\bibinfo{year}{1988}).

\bibitem[{\citenamefont{Bernevig and Zhang}(2006)}]{bernevig}
\bibinfo{author}{\bibfnamefont{B.~A.} \bibnamefont{Bernevig}} \bibnamefont{and}
  \bibinfo{author}{\bibfnamefont{S.-C.} \bibnamefont{Zhang}},
  \bibinfo{journal}{Phys. Rev. lett.} \textbf{\bibinfo{volume}{96}},
  \bibinfo{pages}{106802} (\bibinfo{year}{2006}).

\bibitem[{\citenamefont{Kane and Mele}(2005{\natexlab{a}})}]{kane}
\bibinfo{author}{\bibfnamefont{C.}~\bibnamefont{Kane}} \bibnamefont{and}
  \bibinfo{author}{\bibfnamefont{E.}~\bibnamefont{Mele}},
  \bibinfo{journal}{Phys. Rev. lett.} \textbf{\bibinfo{volume}{95}},
  \bibinfo{pages}{146802} (\bibinfo{year}{2005}{\natexlab{a}}).

\bibitem[{\citenamefont{Hatsugai}(1993)}]{hatsugai}
\bibinfo{author}{\bibfnamefont{Y.}~\bibnamefont{Hatsugai}},
  \bibinfo{journal}{Phys. Rev. B} \textbf{\bibinfo{volume}{48}},
  \bibinfo{pages}{11851} (\bibinfo{year}{1993}).

\bibitem[{\citenamefont{Harper}(1955)}]{harper}
\bibinfo{author}{\bibfnamefont{P.~G.} \bibnamefont{Harper}},
  \bibinfo{journal}{Proc. Phys. Soc. A} \textbf{\bibinfo{volume}{68}},
  \bibinfo{pages}{874} (\bibinfo{year}{1955}).

\bibitem[{\citenamefont{Kane and Mele}(2005{\natexlab{b}})}]{kaneQSH}
\bibinfo{author}{\bibfnamefont{C.}~\bibnamefont{Kane}} \bibnamefont{and}
  \bibinfo{author}{\bibfnamefont{E.}~\bibnamefont{Mele}},
  \bibinfo{journal}{Phys. Rev. lett.} \textbf{\bibinfo{volume}{95}},
  \bibinfo{pages}{226801} (\bibinfo{year}{2005}{\natexlab{b}}).

\bibitem[{\citenamefont{Cohen-Tannoudji
  et~al.}(1992)\citenamefont{Cohen-Tannoudji, Dupont-Roc, and Grynberg}}]{1}
\bibinfo{author}{\bibfnamefont{C.}~\bibnamefont{Cohen-Tannoudji}},
  \bibinfo{author}{\bibfnamefont{J.}~\bibnamefont{Dupont-Roc}},
  \bibnamefont{and} \bibinfo{author}{\bibfnamefont{G.}~\bibnamefont{Grynberg}},
  \emph{\bibinfo{title}{Atom-Photon Interactions}} (\bibinfo{publisher}{Wiley},
  \bibinfo{address}{New York}, \bibinfo{year}{1992}).

\bibitem[{\citenamefont{Murakami et~al.}(2003)\citenamefont{Murakami, Nagaosa,
  and Zhang}}]{2}
\bibinfo{author}{\bibfnamefont{S.}~\bibnamefont{Murakami}},
  \bibinfo{author}{\bibfnamefont{N.}~\bibnamefont{Nagaosa}}, \bibnamefont{and}
  \bibinfo{author}{\bibfnamefont{S.}~\bibnamefont{Zhang}},
  \bibinfo{journal}{Science} \textbf{\bibinfo{volume}{301}},
  \bibinfo{pages}{1348} (\bibinfo{year}{2003}).

\bibitem[{\citenamefont{Gerbier and Dalibard}(2010)}]{dalibard}
\bibinfo{author}{\bibfnamefont{F.}~\bibnamefont{Gerbier}} \bibnamefont{and}
  \bibinfo{author}{\bibfnamefont{J.}~\bibnamefont{Dalibard}},
  \bibinfo{journal}{New Jour.~Phys.} \textbf{\bibinfo{volume}{12}},
  \bibinfo{pages}{033007} (\bibinfo{year}{2010}).

\bibitem[{\citenamefont{Cooper}(2011)}]{opticalflux}
\bibinfo{author}{\bibfnamefont{N.}~\bibnamefont{Cooper}},
  \bibinfo{journal}{Phys. Rev. lett} \textbf{\bibinfo{volume}{106}},
  \bibinfo{pages}{175301} (\bibinfo{year}{2011}).

\bibitem[{\citenamefont{Dalibard et~al.}(2011)\citenamefont{Dalibard, Gerbier,
  Juzeliūnas, and Öhberg§}}]{gauge}
\bibinfo{author}{\bibfnamefont{J.}~\bibnamefont{Dalibard}},
  \bibinfo{author}{\bibfnamefont{F.}~\bibnamefont{Gerbier}},
  \bibinfo{author}{\bibfnamefont{G.}~\bibnamefont{Juzeliūnas}},
  \bibnamefont{and}
  \bibinfo{author}{\bibfnamefont{P.}~\bibnamefont{Öhberg§}},
  \bibinfo{journal}{Rev. Mod. Phys.} \textbf{\bibinfo{volume}{83}},
  \bibinfo{pages}{1523} (\bibinfo{year}{2011}).

\bibitem[{\citenamefont{Jaksch and Zoller}(2005)}]{toolbox}
\bibinfo{author}{\bibfnamefont{D.}~\bibnamefont{Jaksch}} \bibnamefont{and}
  \bibinfo{author}{\bibfnamefont{P.}~\bibnamefont{Zoller}},
  \bibinfo{journal}{Annals of Physics} \textbf{\bibinfo{volume}{315}},
  \bibinfo{pages}{52} (\bibinfo{year}{2005}).

\bibitem[{\citenamefont{Jaksch et~al.}(1998)\citenamefont{Jaksch, Bruder,
  Cirac, Gardiner, and Zoller}}]{zoller}
\bibinfo{author}{\bibfnamefont{D.}~\bibnamefont{Jaksch}},
  \bibinfo{author}{\bibfnamefont{C.}~\bibnamefont{Bruder}},
  \bibinfo{author}{\bibfnamefont{J.~I.} \bibnamefont{Cirac}},
  \bibinfo{author}{\bibfnamefont{C.~W.} \bibnamefont{Gardiner}},
  \bibnamefont{and} \bibinfo{author}{\bibfnamefont{P.}~\bibnamefont{Zoller}},
  \bibinfo{journal}{Phys. Rev. lett} \textbf{\bibinfo{volume}{81}},
  \bibinfo{pages}{3108} (\bibinfo{year}{1998}).

\bibitem[{\citenamefont{Ashcroft and Mermin}(1976)}]{ashcroft}
\bibinfo{author}{\bibfnamefont{N.~W.} \bibnamefont{Ashcroft}} \bibnamefont{and}
  \bibinfo{author}{\bibfnamefont{N.~D.} \bibnamefont{Mermin}},
  \emph{\bibinfo{title}{Solid State Physics}} (\bibinfo{publisher}{Thomson
  Learning}, \bibinfo{address}{Toronto}, \bibinfo{year}{1976}).

\bibitem[{\citenamefont{Kennedy et~al.}(2013)\citenamefont{Kennedy, Siviloglou,
  Miyake, Burton, and Ketterle}}]{spinflip}
\bibinfo{author}{\bibfnamefont{C.~J.} \bibnamefont{Kennedy}},
  \bibinfo{author}{\bibfnamefont{G.~A.} \bibnamefont{Siviloglou}},
  \bibinfo{author}{\bibfnamefont{H.}~\bibnamefont{Miyake}},
  \bibinfo{author}{\bibfnamefont{W.~C.} \bibnamefont{Burton}},
  \bibnamefont{and} \bibinfo{author}{\bibfnamefont{W.}~\bibnamefont{Ketterle}},
  \bibinfo{journal}{Phys. Rev. Lett.} \textbf{\bibinfo{volume}{111}},
  \bibinfo{pages}{225301} (\bibinfo{year}{2013}).

\end{thebibliography}

\end{document}